# Title

Application State Management (ASM) in the Modern Web and Mobile Applications: A Comprehensive Review

# Authors

*Anujkumarsinh Donvir*
*Application Development*
*Wayne, NJ, USA*
anujdonvir@ieee.org
*ORCID: 0009-0006-3243-5780*

*Apeksha Jain*
*Distributed Systems Development*
*Boston, MA, USA*
J.apekshajain@gmail.com
*ORCID: 0009-0005-1007-8033*

*Pradeep Kumar Saraswathi*
*User Interface Architect*
*Fremont, California, USA*
spradeep@ieee.org
*ORCID: 0009-0007-8006-0669*

# Abstract:

The rapid evolution of web and mobile applications has necessitated robust mechanisms for managing application state to ensure consistency, performance, and user-friendliness. This comprehensive review examines the most effective Application State Management (ASM) techniques, categorized into Local State Management, State Management Libraries, and Server-Side State Management. By analyzing popular front end frameworks the study delves into local state management mechanisms. It also evaluates the state of front end management libraries, highlighting their implementations, benefits, and limitations. Server-side state management techniques, particularly caching, are discussed for their roles in enhancing data retrieval efficiency. This paper offers actionable insights for developers to build scalable, responsive applications, aiming to bridge the gap between theoretical knowledge and practical application. This study's critical analysis and recommendations aim to guide future research and development in ASM, contributing to the advancement of modern application architecture.

# Keywords:

Application State Management (ASM), React State Management, Redux vs MobX, Server-Side State Management, Modern Web Application Development.

# I Introduction:

The landscape of modern web and mobile applications is increasingly expanding with countless apps being built continuously. These apps are highly dynamic and interactive in nature, which present strong challenges. Applications need to stay consistent in their behavior, need to be user friendly and perform blazing fast. Application State Management (ASM) has fast become a go-to mechanism for achieving these needs of modern applications.[1] ASM at its core is management of data and user interactions throughout the application's lifecycle, which defines the stages of initialization, execution, and termination of an application. ASM  streamlines an application's data, and mutations (changes) to this data making the data predictable at various stages of the lifecycle. This predictability allows all parts of an application to access this data with tools provided by various ASM techniques when required and present it to the user, making the entire application consistent and responsive.

Furthermore, the Single Responsibility Principle (SRP) [22], introduced by Robert Martin in 1972, a key design pattern for web and mobile applications states that software components should have only one reason to change. SRP principle leads to creating scalable and maintainable User Interface (UI) and Application Programming Interfaces (APIs), by ensuring each component carries out one responsibility only, a key need for modern applications. ASM allows creation of software components which display as well react to small amounts of information in the grand scheme of entire application data. These benefits provide great opportunities to carry out ASM at client ( web browser or mobile app) as well as server side by making application architecture robust and cohesive.

This article aims to comprehensively review various ASM techniques by broadly categorizing them in three categories of Local State Management, State Management Libraries, and Server-Side State Management. These categories have a large list of tools and techniques which suits different aspects of application development and use. First, the paper explores local state management in popular user interface development frameworks such useState in React, component properties in Angular and data properties in Vue. Next, it analyzes modern state management libraries such as Redux, MobX, NgRx, Ngxs, Vuex, Recoil, and Zustand [2]. Following that, the paper focuses on synthesizing use cases, benefits and importance of Server-Side State Management using techniques like View State, Session State and Application State. This paper carries out critical analysis of  effectiveness, adoption, current trends and future scope of these ASM techniques and recommends key use cases, benefits and disadvantages of using each of these. This should allow readers of the paper with actionable insights to build scalable, user friendly and responsive web and mobile applications.

## II Methodology:

To conduct this comprehensive review, a multifaceted approach was adopted. Extensive literature review was conducted, recent trend surveys conducted by leading organizations were studied, thought leadership pieces published by technology industry experts were examined and expert opinions from experienced application development experts were taken. While evaluating each option, typical use cases for the options have also been documented. This approach has resulted in recommendations which are practical and forward looking in nature for application state management (ASM) techniques, moving beyond barriers of just theoretical assessments.

For evaluation of ASM techniques which focus on the User Interface (UI) of an application, React, Angular and Vue frameworks have been considered. Underlying reason for this choice is due to the fact that React, Angular and Vue are leading frameworks used while building the UI of modern applications. Usage of other frameworks has not reached comparable levels of significance as per the survey conducted by The State of JS. [3]

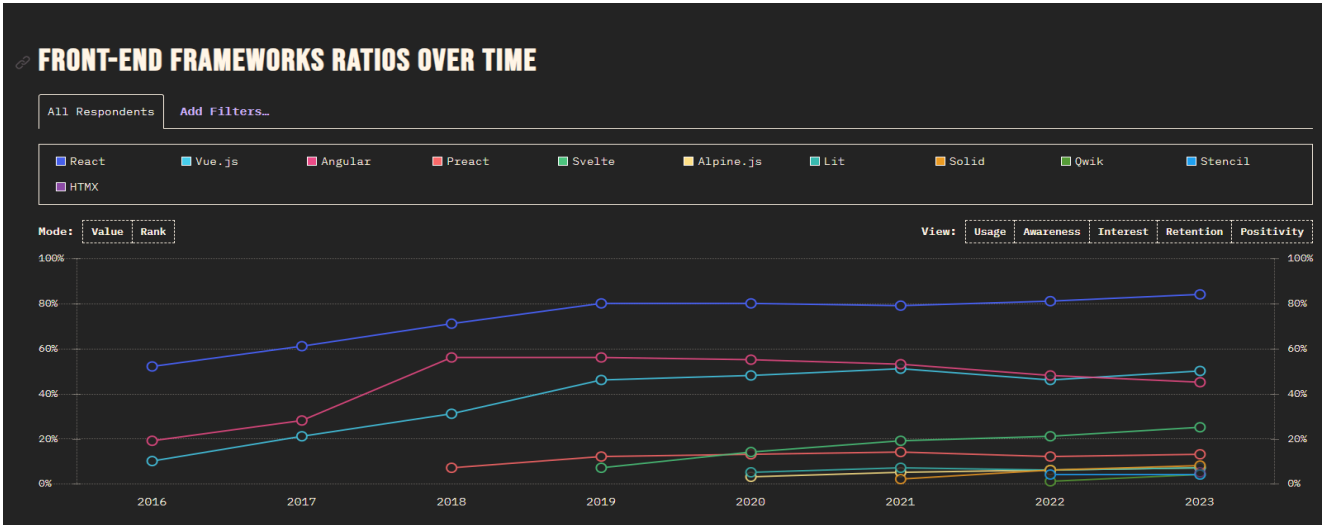

State of JS Survey 2023

Similarly, while evaluating server-side Application State Management (ASM) techniques, this paper aims to highlight the role of major databases utilized for caching, such as Redis and Memcached, in ASM. According to the 2023 Stack Overflow survey [4], Redis is one of the leading storage databases used for caching. Notably, while other databases rank higher in overall popularity, their primary use cases do not typically involve caching mechanisms. This paper also examines traditional databases such as SQL and other key-value pair databases for ASM. Additionally, the paper delves into server-side ASM technologies, exploring technologies such as the Blazor server.

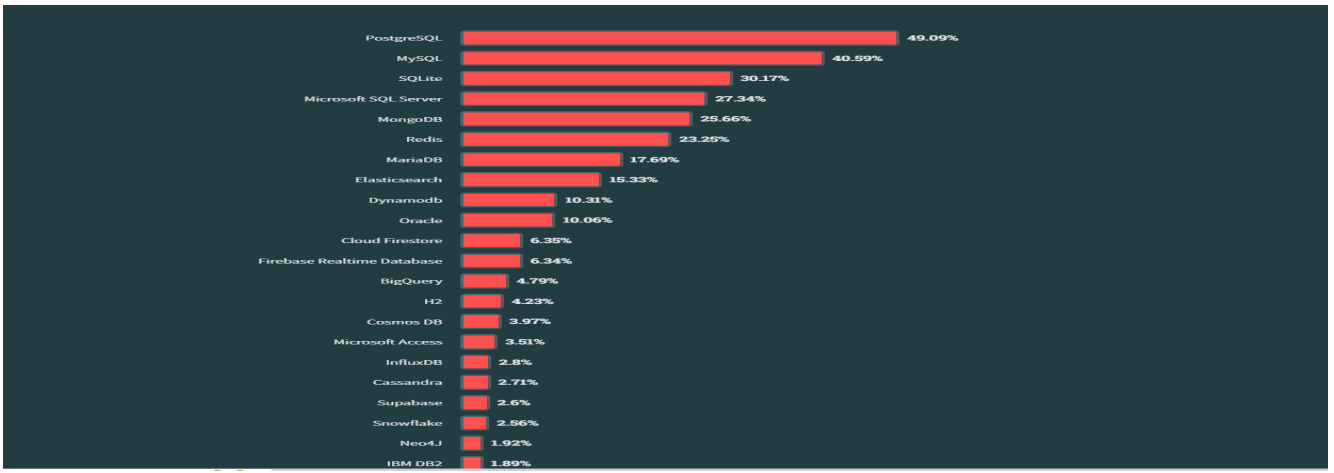

StackOverflow most popular technologies for year 2023

When analyzing these techniques expert opinions were taken. These experts have worked with various ASM tools and frameworks, often for multiple projects. These experts include developers and architects for user interface, backend, APIs and databases. They have shared their wins and losses with each tool and provided detailed opinions on multiple techniques based on their focus area. Furthermore, key statistics such as downloads on popular package managers, and stars and forks on GITHUB to understand market adoption trends were captured.

**2.1 Local State Management:** Local State Management plays a crucial role in handling data which is related to one component or simply one section of the user interface of the application. This data is not required to be shared with other parts of the application, thus limiting data propagation and improving performance of the application. Important uses cases of Local State Management can be:

**Use Cases:**
- **Form Handling:**
  *Scenario*: When handling form inputs inside an application, generally user input data doesn't need to be shared across various components in the application.
  *Use*: When creating a user sign up form, information such as first name, last name, email id and password needs to be captured. This data may need validation such as password being specific length and containing special characters, and they need to be submitted to the server directly instead of being passed around in the app.
- **Presentation Logic Change:**
  *Scenario*: A component may already have all the necessary information it needs to show to the user, and it may offer multiple ways the user can view the data on their screen. Local state management becomes a valuable tool in such scenarios.
  *Use:* A component which shows price over a period of time for any stock, user may want to see data in table format or in a graph format. In this case, the component already has necessary stock data, and can allow the user to toggle a button to show graph vs table by capturing toggle button's state in a local state variable. Users may want to reorder data by changing sort to see latest stock prices first or old stock prices first, and this use case just like the toggle button makes a good use of local state variables.
- **Handling Notification and Updates:**
  *Scenario:* Real time data such as notifications do not need to be perceived, but just displayed onto the User Interface immediately. Local state variables allow relaying such information to the User Interface without any need to save notification application wide.
  *Use:* An application that allows large file upload to its users. Large file uploads can take a lot of time. It is not a good practice to block users from accessing other parts of the application, and at the same time the user must be informed of successful or failed file upload. An asynchronous file upload mechanism to upload the file and then a local state variable to trigger notification with file upload status helps implement this logic efficiently.

**2.1.1 Local State Management in React, Angular and Vue:** This examination explores the options available to perform ASM using local state management with an example of async file upload.

**2.1.1.1 React:** React has two types of components - functional components and class components. [5] In React class components local state management is done with this.state variable, while in React functional components the state management is done using useState hook. This can be better understood with the code snippet presented ahead.
In the presented code snippet, a local state variable named uploadProgress is defined. Its value is dynamically modified upon successful file upload using the setUploadProgress method. The useState hook of React used for initialization as well as declaration of the update method. React hooks are a fundamental mechanism managing component-level state inside React.[20]

```jsx
import React, { useState } from 'react';

const FileUploadComponent = () => {
  const [file, setFile] = useState(null);
  const [uploadProgress, setUploadProgress] = useState(0);

  const handleFileChange = (e) => setFile(e.target.files[0]);

  const handleUpload = async () => {
    const formData = new FormData();
    formData.append('file', file);

    try {
      const response = await fetch('https://api.example.com/upload', {
        method: 'POST',
        body: formData,
        onUploadProgress: (progressEvent) =>
setUploadProgress(Math.round((progressEvent.loaded / progressEvent.total) * 100)),
      });

      console.log('Upload complete:', response);
    } catch (error) {
      console.error('Upload failed:', error);
    }
  };

  return (
    <div>
      <input type="file" onChange={handleFileChange} />
      <button onClick={handleUpload}>Upload File</button>
      {uploadProgress > 0 && <p>Upload Progress: {uploadProgress}%</p>}
    </div>
  );
};

export default FileUploadComponent;
```

**2.1.1.2 Angular:** Components in Angular are JavaScript classes, which allows declaration of state variables right at the at class code initiation. State variables can be assigned initial value during declaration or in the constructor method of the class. Another good place to initialize their value is in the ngOnInit lifecycle method provided by Angular framework. [6] To illustrate the concept, the same file upload mechanism has been implemented in Angular.

In the code snippet ahead, a local state variable named uploadProgress which tracks status of file upload in percentage value has been created. Variable has been initialized to 0 while declaration. When doing the upload of file in handleUpload in method, a method named onUploadProgress is received as callback [21] (function that is passed as an argument to another function and is executed after a particular event or condition occurs), which is used to update value of local state variable uploadProgress.

```
// Component code
import { Component } from '@angular/core';

@Component({
  selector: 'app-file-upload',
  templateUrl: './file-upload.component.html',
  styleUrls: ['./file-upload.component.css']
})
export class FileUploadComponent {
  file: File | null = null;
  uploadProgress: number = 0;

  handleFileChange(event: any) {
    this.file = event.target.files[0];
  }

  async handleUpload() {
    const formData = new FormData();
    formData.append('file', this.file as File);

    try {
      const response = await fetch('https://api.example.com/upload', {
        method: 'POST',
        body: formData,
        onUploadProgress: (progressEvent) => {
          this.uploadProgress = Math.round((progressEvent.loaded / progressEvent.total) * 100);
        },
      });

      console.log('Upload complete:', response);
    } catch (error) {
      console.error('Upload failed:', error);
    }
  }
}
```

```
<!-- html -->
<div>
 <input type="file" (change)="handleFileChange($event)" />
```

```
  <button (click)="handleUpload()">Upload File</button>
  <p *ngIf="uploadProgress > 0">Upload Progress: {{ uploadProgress
}}%</p>
</div>
```

**2.1.1.3 Vue:** Like Angular, in Vue also components are JavaScript classes allowing creation of class variables that act as local state variables. [7] These variables can mutate (change value) via callback methods. Demonstration of local state management with same example of file upload functionality is provided ahead:

In Vue, the data method lets you define initial state variables and data properties, at the same time making them reactive ( updating the value of these variables in the updated view on the user interface). In the code, uploadProgress local state variable is defined inside the data method and initialized to value 0. Upon receiving details from onUploadProgress callback, the value of uploadProgress is updated. This change gets reflected on the user interface on the progress bar.

```
<template>
  <div>
    <input type="file" @change="handleFileChange" />
    <button @click="handleUpload">Upload File</button>
    <p v-if="uploadProgress > 0">Upload Progress: {{ uploadProgress
}}%</p>
  </div>
</template>

<script>
export default {
  data() {
    return {
      file: null,
      uploadProgress: 0,
    };
  },
  methods: {
    handleFileChange(event) {
      this.file = event.target.files[0];
    },
    async handleUpload() {
      const formData = new FormData();
      formData.append('file', this.file);

      try {
        const response = await fetch('https://api.example.com/upload', {
          method: 'POST',
          body: formData,
          onUploadProgress: (progressEvent) => {
```

```
        this.$set(this, 'uploadProgress',
Math.round((progressEvent.loaded / progressEvent.total) * 100));
        },
      });

      console.log('Upload complete:', response);
    } catch (error) {
      console.error('Upload failed:', error);
    }
   },
  },
};
</script>
```

**2.1.2 Review of Local State Management:**
- Local State Management can help streamline the application and boost its performance by efficiently handling data that doesn't need sharing across various components of the applications.
- Local State Management can be applicable to a wide range of use cases such as form handling, asynchronous operations and presentation change limited to scope of a particular component.
- Modern User Interface frameworks React, Angular and Vue have their own implementation of local state management.
    - React relies on state variables in class components and hooks in functional components. [8]
    - Angular utilizes class properties to implement class properties with multiple options for initialization of these variables.
    - Vue uses special method data() to implement and return local state variables.
    - Angular uses change detection (a special process which runs in background by Angular framework) to update the User Interface when state variables change while React and Vue automatically trigger the update to User Interface when state variables get updated.
- Local State Management techniques are tides to individual frameworks, making it hard to create interoperability between different applications.
- Local State Management can cause tight coupling between different parts of the application if not used correctly, as too much data needs to be passed around between components. This makes components interdependent.

**2.2 State Management Libraries:** State management libraries allow efficient use of the data which need to be shared across multiple parts of the application. This data needs to be consistent when accessed in different sections, which requires a robust mechanism for reliably updating the data.
**Use Cases:**
- **Global State Management:**
  *Scenario*: Certain information needs to be shared across many components in an application, and updates to this information must stay consistent inside the entire application.
  *Use*: In an e-commerce application, information about items added to cart must stay consistent when an user is on cart page, checkout page, item description page or any other page. Furthermore, the quantity of items can be edited at multiple parts (components) in the application, and this change should get reflected in all other places referring to the data.

- **Cross-Component Communication:**
  *Scenario*: Sharing of data between two or more components can be streamlined using state management libraries, and these components can have parent-child or sibling relations.
  *Use*: Implementation of this use case is possible in many kinds of scenarios across inter component communication. For example, in an insurance application users can try to find the best rate by changing various clauses such as amount of coverage, deductible, address etc. All theses information, which can be changed on a fly should be actively passed to the component calculating insurance premium rates.
- **Undo and Redo Functionality:**
  *Scenario*: An application may allow its users to revert an action done or cancel this revert action also.
  *Use*: An education application may have an interactive medium for its students to solve math problems, and during the course of the solution, students may undo a specific step they took and later come to understand they made a mistake. They may choose to redo the step again. State Management Libraries easily allow the users to travel through the time and reapply any previous actions.

**2.2.1 State Management Libraries for React, Angular and Vue:** React ecosystem is very loosely opinionated about how any application should be built using React, while Angular and Vue are more structured frameworks for building applications. This has allowed for multiple libraries to emerge for the React ecosystem, and this paper focuses on Redux, Recoil, Zustand and Mobx as State Management Libraries for React. For Angular, NgRx has been analyzed , while for Vue, the Vuex has been analyzed.

**2.2.2.1 Redux for React:** Redux has one single data source referred to as the store and it should be shared across the entire application, making the data source a single source of truth. The Data in the Store can not be modified (mutated), and when change of data inside the store is warranted, a new copy of the store data is created. This new copy is achieved through functions called reducers. Reducers change the data based events called actions. Actions are plain JavaScript objects that describe what happened, and how store data should be updated. When an action is dispatched, the reducer function takes the current state and the action as arguments, and returns a new state. Optionally Redux allows middleware functions to do extra work, referred to as side effects. One example of side effects is calling an API to load the data into the store. [9][19]

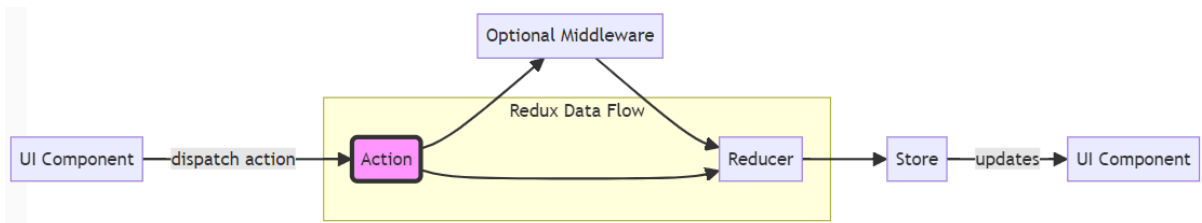

Redux Flow

Below code shows an example of how Redux can be used to implement a simple account management part in a banking / financial application. Code demonstrates viewing balance, and actions such as incrementing and decrementing the balances.

Inside code, actions.js defines two actions for incrementing and decrementing the balance. These actions get acted upon by the balanceReducer. Moreover, the balanceReducer also defines the initial balance to be 0. Finally, this reducer is taken as an argument when creating the store.

```javascript
// actions.js
export const INCREMENT = 'INCREMENT';
```

```javascript
export const DECREMENT = 'DECREMENT';

export const incrementBalance = (amount) => ({
  type: INCREMENT,
  payload: amount,
});

export const decrementBalance = (amount) => ({
  type: DECREMENT,
  payload: amount,
});
// reducer.js
import { INCREMENT, DECREMENT } from './actions';

const initialState = {
  balance: 0,
};

const balanceReducer = (state = initialState, action) => {
  switch (action.type) {
    case INCREMENT:
      return {
        ...state,
        balance: state.balance + action.payload,
      };
    case DECREMENT:
      return {
        ...state,
        balance: state.balance - action.payload,
      };
    default:
      return state;
  }
};

export default balanceReducer;
// store.js
import { createStore } from 'redux';
import balanceReducer from './reducer';

const store = createStore(balanceReducer);

export default store;
```

**2.2.2.2 Recoil for React:** Recoil is another popular library that takes a more fine-grained and flexible approach to implementing state management. Application data is broken into smaller units, referred to as atoms, compared to having a single source for the entire data as in the case of Redux. These atoms can independently be read and updated by components; only components that have subscribed to an atom get re-rendered upon an update to the atom. Selectors are special functions in the Recoil library that allow access to an atom's data (the state), and multiple selectors can be used together to get the derived state for complex data logic. Selectors also handle side effects such as calling APIs. [10]

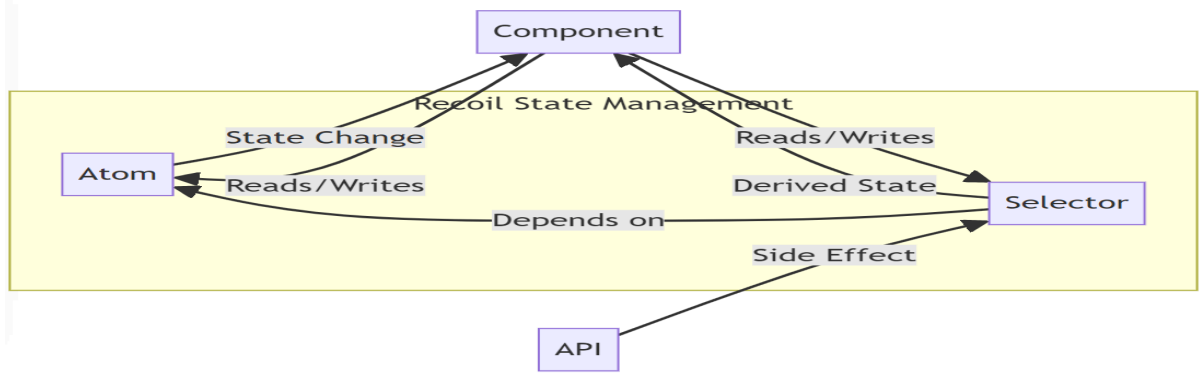

RECOIL Data Flow

Code below provides a basic outline for implementing a similar feature of the banking system, where account balance can be incremented and decremented, a similar functionality which was demonstrated in section about Redux.

A balanceState atom is defined with initial balance of 0, and this atom is used to create balanceSelector. Creation of an application-wide data store is not needed. Individual components of the React application can access as well modify data inside this atom using the useRecoilState hook provided by the Recoil library.

```
// state.js
import { atom, selector } from 'recoil';

// Atom to store the balance
export const balanceState = atom({
  key: 'balanceState',
  default: 0,
});

// Selectors can be added if derived state is needed
export const balanceSelector = selector({
  key: 'balanceSelector',
  get: ({ get }) => {
    const balance = get(balanceState);
    return balance;
  },
});
// BankingApp.js
import React from 'react';
import { RecoilRoot, useRecoilState, useRecoilValue } from 'recoil';
import { balanceState } from './state';
```

```
const BankingApp = () => {
  const [balance, setBalance] = useRecoilState(balanceState);

  const handleIncrement = () => {
    setBalance(balance + 10); // Increment balance by 10
  };

  const handleDecrement = () => {
    setBalance(balance - 5); // Decrement balance by 5
  };

  return (
    <div>
      <h1>Banking App</h1>
      <h2>Balance: ${balance}</h2>
      <button onClick={handleIncrement}>Increment Balance</button>
      <button onClick={handleDecrement}>Decrement Balance</button>
    </div>
  );
};
```

**2.2.2.3 Zustand for React:** Zustand offers simplified APIs to implement Global State Management for the entire application. Importantly Zustand cuts down the boilerplate code (initial setup code) needed to implement state management. It relies on React hooks to access and manage state data. Zustand allows the creation of a single store that can be sliced and diced for consumption inside React components, offering ease of implementation. Zustand allows for middleware integrations to handle side effects. [11]

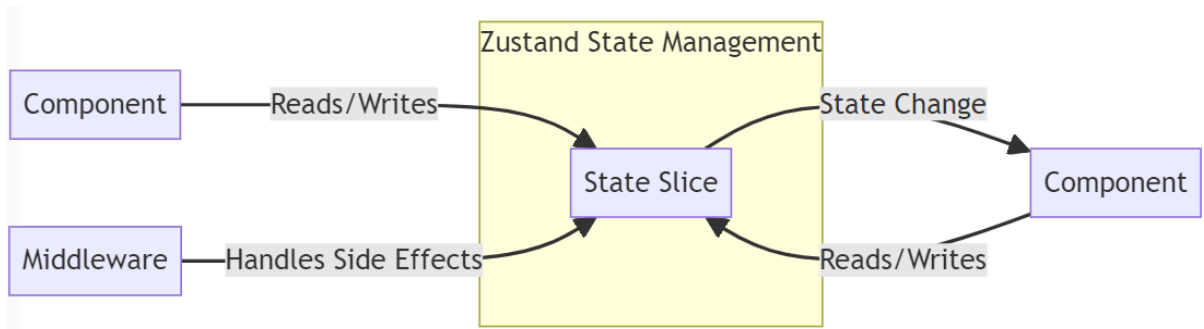

Zustand Data Flow

Basic implementation of the same balance increase and decrease operation for the piece of a banking application, core features are kept the same as before like Redux and Recoil for easy comparison .

A new hook useStore is created using Zustand's create API. During creation, the initial state ( in this case initial balance) is set at 0 and function definitions for incrementing and decrementing this balance are provided. The BankingApp component subscribes to the state value, and methods using created useStore hook.

```
// store.js
import create from 'zustand';
```

```javascript
const useStore = create((set) => ({
  balance: 0,
  incrementBalance: (amount) => set((state) => ({ balance: state.balance + amount })),
  decrementBalance: (amount) => set((state) => ({ balance: state.balance - amount })),
}));

export default useStore;
// BankingApp.js
import React from 'react';
import useStore from './store';

const BankingApp = () => {
  const balance = useStore((state) => state.balance);
  const incrementBalance = useStore((state) => state.incrementBalance);
  const decrementBalance = useStore((state) => state.decrementBalance);

  const handleIncrement = () => {
    incrementBalance(10); // Increment balance by 10
  };

  const handleDecrement = () => {
    decrementBalance(5); // Decrement balance by 5
  };

  return (
    <div>
      <h1>Banking App</h1>
      <h2>Balance: ${balance}</h2>
      <button onClick={handleIncrement}>Increment Balance</button>
      <button onClick={handleDecrement}>Decrement Balance</button>
    </div>
  );
};

export default BankingApp;
```

**2.2.2.4 MobX for React:** MobX has gained quick popularity in the React ecosystem as a state management library due to its robust and flexible approach to allowing state management across the entire application. MobX relies on reactive programming (Programming paradigm for asynchronous data streams and event handling) [23] and decorators (Functions modifying behavior of classes or methods at runtime) for implementing the state, effectively cutting down boilerplate code. MobX breaks the entire state into small observables, allowing for easy consumption of state data via subscriptions. State can be modified using actions by individual components. MobX offers functions called Reactions to manage side effects, such as handling asynchronous activities. Reactions are functions that automatically run whenever the observable state they depend on changes. [12]

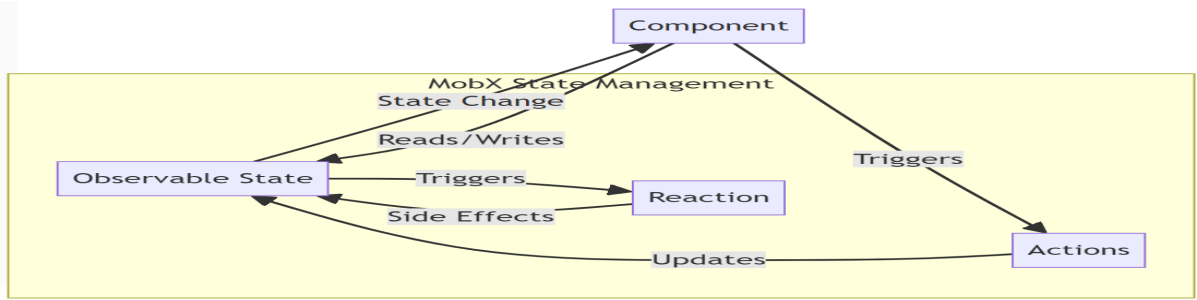

MobX Dat Flow

Below sample code implements balance increase and decrease functionality for an account in a banking application as demonstrated before for other frameworks.

A new store class is created using the makeAutoObservable API from MobX inside the class constructor (a function that runs when a class instance is created). After that, an instance of this Store class is created and exported. A new component is created using the observer API offered by MobX. This component has access to the value of the balance as well as methods to increment and decrement the balance.

```
// store.js
import { makeAutoObservable } from 'mobx';

class Store {
  balance = 0;

  constructor() {
    makeAutoObservable(this);
  }

  increment(amount) {
    this.balance += amount;
  }

  decrement(amount) {
    this.balance -= amount;
  }
}

const store = new Store();
export default store;
// App.js
import React from 'react';
import { observer } from 'mobx-react';
import store from './store';

const BankingApp = observer(() => {
```

```
  const handleIncrement = () => {
    store.increment(10); // Increment balance by 10
  };

  const handleDecrement = () => {
    store.decrement(5); // Decrement balance by 5
  };

  return (
    <div>
      <h1>Banking App</h1>
      <h2>Balance: ${store.balance}</h2>
      <button onClick={handleIncrement}>Increment Balance</button>
      <button onClick={handleDecrement}>Decrement Balance</button>
    </div>
  );
});

export default BankingApp;
```

**2.2.2.5 NgRx for Angular:** NgRx combines the power of RxJS (library used to implement Reactive Programming in JavaScript) with the principles of Redux architecture to create a framework for state management in Angular applications. It has a single store for data across the entire application which can not be mutated. Additionally, it comes with a clean way to manage side effects. It allows direct integration with Angular Router (a module which handles routing between pages in an application), and boasts robust developer tooling for efficient development. [13]

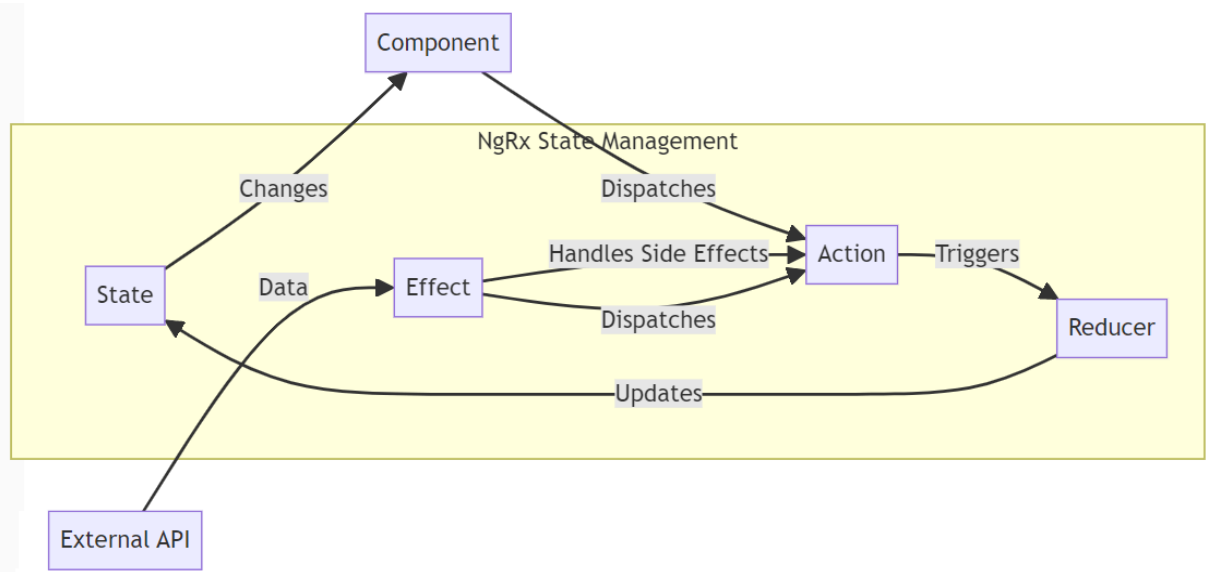

NgRx Data Flow

Below code provides basic implementation for NgRx store in an Angular based application.
Actions for incrementing and decrementing the balance are defined inside balance.actions.ts file. These actions are used to create balanceReducer which holds logic for initial balance as well. Selectors are created to consume the state, and finally the store is initiated inside the app module. Any Angular component within the app module has access to the store.

```ts
// balance.actions.ts
import { createAction, props } from '@ngrx/store';

export const incrementBalance = createAction(
  '[Balance] Increment',
  props<{ amount: number }>()
);

export const decrementBalance = createAction(
  '[Balance] Decrement',
  props<{ amount: number }>()
);
// balance.reducer.ts
import { createReducer, on } from '@ngrx/store';
import { incrementBalance, decrementBalance } from './balance.actions';

export interface BalanceState {
  balance: number;
}

export const initialState: BalanceState = {
  balance: 0,
};

export const balanceReducer = createReducer(
  initialState,
  on(incrementBalance, (state, { amount }) => ({
    ...state,
    balance: state.balance + amount,
  })),
  on(decrementBalance, (state, { amount }) => ({
    ...state,
    balance: state.balance - amount,
  }))
);
// balance.selectors.ts
import { createSelector, createFeatureSelector } from '@ngrx/store';
import { BalanceState } from './balance.reducer';

export const selectBalanceState =
createFeatureSelector<BalanceState>('balance');

export const selectBalance = createSelector(
```

```
    selectBalanceState,
    (state: BalanceState) => state.balance
);
// app.module.ts
import { NgModule } from '@angular/core';
import { BrowserModule } from '@angular/platform-browser';
import { StoreModule } from '@ngrx/store';
import { balanceReducer } from './balance.reducer';
import { AppComponent } from './app.component';

@NgModule({
  declarations: [
    AppComponent
  ],
  imports: [
    BrowserModule,
    StoreModule.forRoot({ balance: balanceReducer })
  ],
  providers: [],
  bootstrap: [AppComponent]
})
export class AppModule { }
```

**2.2.2.6 Vuex for Vue:** Vuex is a library for state management for applications built on the Vue framework. Similar to NgRx, it relies on Redux and Reactive programming to achieve immutable data store state and has extensive developer tooling. It comes with a mechanism to manage side effects cleanly and has strong integration with the Vue ecosystem. [14]

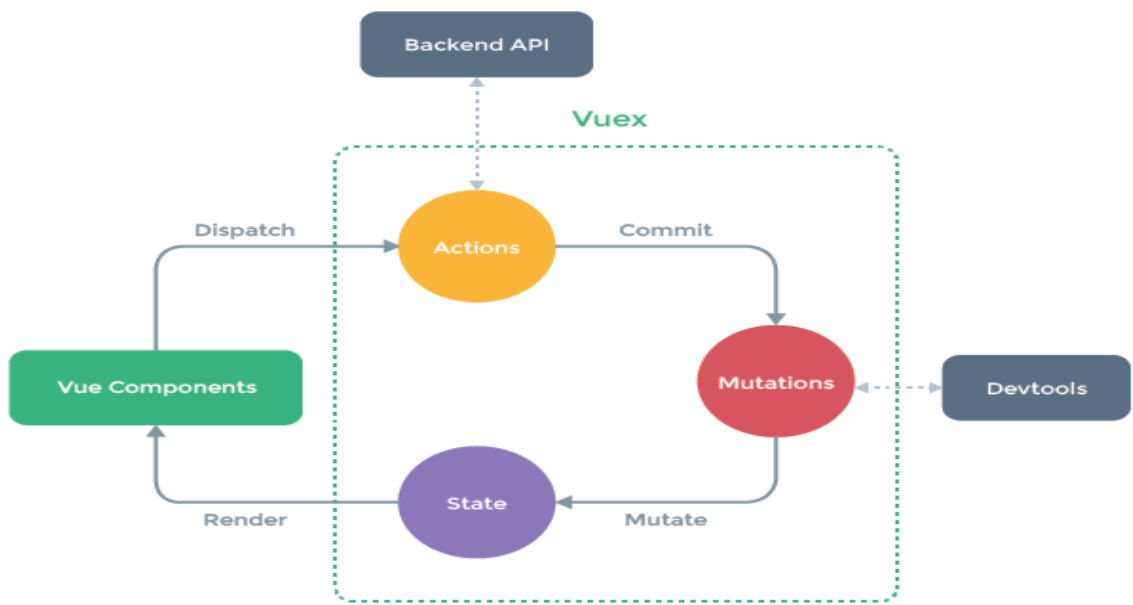

Vuex Data Flow

Below code gives an example for Vuex state management implementation. It implements the same scenario of balance increase and decrease functionality for a banking system.

A new store which handles state management is initiated with Vuex's Store API, and the state has a balance variable with a starting value of 0. The definition includes an object known as mutations which implements logic of balance increase and decrease. Actions object creates definitions which are used to commit changes to store via the mutations. Getters are defined to allow components to access the values in the store.

```js
// store.js
import Vue from 'vue';
import Vuex from 'vuex';

Vue.use(Vuex);

export default new Vuex.Store({
  state: {
    balance: 0
  },
  mutations: {
    incrementBalance(state, amount) {
      state.balance += amount;
    },
    decrementBalance(state, amount) {
      state.balance -= amount;
    }
  },
  actions: {
    incrementBalance({ commit }, amount) {
      commit('incrementBalance', amount);
    },
    decrementBalance({ commit }, amount) {
      commit('decrementBalance', amount);
    }
  },
  getters: {
    balance: state => state.balance
  }
});
```

**2.2.3 Review of State Management Libraries:**
- Redux is the original inspiration for many of the modern state management libraries.
    - Redux has a store that doesn't mutate. This allows debugging and tracking past actions and state of the store seamless.
    - The setup of Redux involves a fair amount of boilerplate code. This can be overcome with 3rd party utilities such as Redux Toolkit.
    - Unopinated approach of Redux allows different teams to implement Redux stores differently even within the same organization, which can lead to code maintenance overhead.

- - Redux needs support of 3rd party tools to implement developer tooling and middleware.
- Despite the need of external tools for better architecture of the application, Redux is ruling when it comes to state management libraries for React. This is because of its battle proven track record and strong community support. Downloads from the npm repository for Redux are higher than any other library.

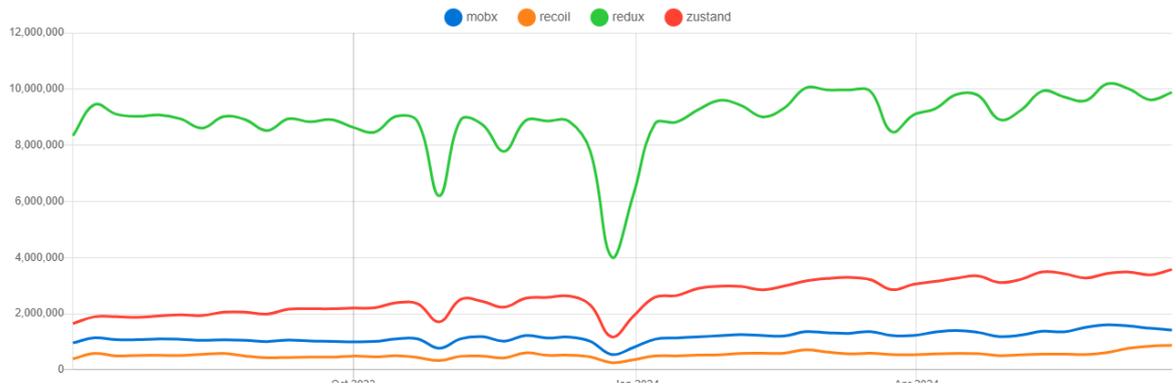
NPM Download Trends

- Ease of implementation has allowed Zustand to gain popularity when it comes to State Management for React Applications.
  - Due to minimal boilerplate code needed, it is easy to get up and running with Zustand.
  - Simple API of Zustand allows for creation of flexible state management with small slices of state. This has performance benefits, as only part of the rendered application which depends on a given slice needs updating instead of the entire application when a part of data in the slice gets updated.
  - It uses React hooks to integrate state management with components, which fits well with the current recommended approach of functional component development in React.
- Recoil shines when fine grained control over state is necessary.
  - Breaking state into smaller atoms allows easy management of state by making pieces of state independent of each other.
  - Selectors in Recoil allow seamless side effect handling and creating of derived state with mix and match of selectors.
  - Recoil, like Zustand, uses hook based API for integration with React.
  - Recoil has robust developer tooling allowing for ease of development and debugging.
- MobX has quickly gained fame with its React ecosystem.
  - MobX brings principles of Reactive Programming, and thanks to its decorators based implementation enjoy low boilerplate code for initial setup.
  - Reactive Programming allows it to scale seamlessly as application grows.
  - MobX offers DevTools for debugging and tracking state changes, thus allowing seamless development experiments.
  - Developers may face a steep learning curve when first interacting with MobX due to additional learnings and complexity involved for Reactive programming.
- NgRx has become a goto solution for state management for Angular application.
  - NgRx uses the same principles of Redux such as immutable state, reducers and actions.
  - It integrates strongly with Angular framework and provides robust developer tooling.
  - It utilizes RxJS for reactive programming, making it suitable for handling asynchronous operations.
  - NgRx offers built-in side effect handling.

- NgRx is not a junior developer friendly library due to reactive programming, and the need for a lot of boilerplate code.
- VueX is the official state management library for the Vue framework.
  - Vuex maintains a single source of truth for state, which can be changed with mutations and actions.
  - It allows for getters and derived states for complex logic
  - Vuex generally involves limited boilerplate code, and offers ease of setup.
  - Vuex's integration with Vue DevTools makes development with Vue seamless.

**2.3 Server Side Management:**
Server side state management is a methodology to manage the state of an application at the server side. The methodology plays a crucial role in improving the application performance and consistency. The client side state management techniques as discussed in the previous section of the article along with the server side state management techniques provides an end to end application consistency, synchronization between client and server side states, improved latency, better user experience, and resiliency.

**Use Cases:**
- **Form Data:**
  *Scenario*: Preserving user input in a multi-step form to ensure data is not lost when navigating between steps.
  *Use*: In a registration application, user information collected in a multi-step form (such as personal details, address, payment information) needs to be preserved as the user navigates between different steps of the form. This ensures that users do not lose their input data if they need to go back to a previous step to make changes or if they refresh the page.
- **User Authentication:**
  *Scenario*: Storing a user's login credentials and session token to maintain their logged-in status across different pages.
  *Use*: In a web application, when a user logs in, their authentication credentials and session token need to be stored to prevent relogins as they navigate through different pages. This ensures that the user does not need to provide credentials every time they access a new page within the application. For example, in a banking application, once the user logs in, they should remain authenticated while navigating between the account summary, transaction history, and transfer funds pages.
- **E-commerce Shopping Cart:**
  *Scenario*: Maintaining a consistent view of the items in a user's shopping cart across all pages and components within the application.
  *Use:* In an e-commerce application, the items a user adds to their shopping cart need to be consistent and updated across various pages, such as the product listing page, product details page, cart page, and checkout page. When the user updates the quantity or removes items from the cart on any page, these changes should be reflected immediately across all other components displaying the cart information to ensure a seamless shopping experience.

**2.3.1 Server-Side State Management Techniques:**
Server side state management can be facilitated via techniques described ahead.

**2.3.1.1 View State:** [15] The view state is used to maintain the state of a web page between postbacks. The scope of this state is limited to a web page and it is not shared across pages. The view state management is achieved by sending the page content back and forth between the HTTP requests. The state is included in the page's HTML, hence, ViewState can increase the HTML page size, expanding to large encoded blocks of text, perhaps in Megabytes. The approach is viable for sharing small amounts of data. Since this approach has limitations for the size of data it can support, another approach offered by Blazor Server for maintaining view state at the server side is discussed in the following section.

**2.3.1.2 Session State**: [15] [16] The session state is used to store user specific data for a user's session. The scope of this state spans across multiple pages and multiple requests accessed in the same session. This state is effective for storing user authentication information, user preferences, and session state such as shopping cart contents. The data can be stored on a state server, in-memory, or persisted on a data store depending on the persistence and scalability requirements of an application. These approaches are discussed in the following section.

Example code [15]: This is a pseudocode where a Session object is used to store and retrieve the data for a specific user. It provides a dictionary-like interface.

```
// Storing data in a session state
Session["username"] = "JohnDoe";

// Retrieving data from session state
String storedUsername = (string)Session["username"];
Console.WriteLine(storedUsername); //Output: JohnDoe

// Updating data in session state
Session["username"] = "JaneSmith";
Console.WriteLine(Session["username"].cart); //Output: user cart info

// Removing data from session state
Session.Remove("username")
```

**2.3.1.3 Application State**: [15] [16] The application state is used to store application level data that is common for all the users accessing the application. This data is available throughout the lifespan of the application to all the users of the application. It is used to store application configuration settings, health of the application, metrics of the application, or any other shared data that needs to be accessed by all users of the application. The application data can be stored on a state server, in-memory, or persisted on a data store, however, it is recommended to persist the data on a data store so that it can be retrieved between server restarts and is also effective in multi-server deployments.

Example code [16]: This is a pseudocode where the Application object is used to store and retrieve application specific data. It provides a dictionary-like interface.

```
// Storing data in application state
Application["visitorCount"] = 100;
Application["apiSuccessRate"] = 99;
```

```
// Retrieving data from application state
Int visitorCount = (int)Application["visitorCount"];
Float apiSuccessRate = (float)Application["apiSuccessRate"];

Console.WriteLine(visitorCount); //Output: 100
Console.WriteLine(apiSuccessRate); //Output: 99

// Updating data in application state
Application["visitorCount"] = visitorCount + 1;
Application["apiSuccessRate"] = apiSuccessRate + 0.1;

// Removing data from application state
Application.Remove("visitorCount")
```

The above code is an example of storing and retrieving application data. However, the Application object does not persist data across multiple servers. The state of the Application object is lost between restarts.

**2.3.2 Technologies for server side state management:**

**2.3.2.1 Blazor server**: [16] The Blazor server is useful for storing a View State. The server provides functionality to the apps for maintaining their state, called a *circuit* for an ongoing connection. The state is stored in the server memory while the connection is active. The *circuit* data is disposed when the user navigates away from the page. The user's circuit is useful for storing the state of [17]:
   A. The UI being rendered i.e. the hierarchy of component instances and their most recent render output.
   B. The values of the fields and properties in component instances.
   C. Data held in the Dependency Injection (DI) service instances that are in scope of the circuit.

Since, circuit is stored in the server's memory and not persisted in a datastore, it requires maintaining sticky sessions to ensure that all requests from the same browser are routed to the same server. If the request goes to a different server, the state is lost [16].

The technique is effective in restoring the state when the user experiences a temporary network connection loss. The Blazor server will provide the last circuit state stored in the server's memory on reconnection. The last state does not need to be rebuilt in this scenario. However, the technique cannot retain circuit across server restarts or in multi-server deployments. In such a case, persisting the state in a data store will be a viable solution.

**2.3.2.2 Caching**: Caching is another mechanism for storing view state, session state, and application state. The selection of technology depends on the application requirements.
   A. Redis [18] - Redis provides an in-memory data store, it stores the data in memory and also persists the data in a data store. Since the client interfaces with Redis memory for read and write operations, it gives a very high performance.

B. Memcache [19] - Memcache is an in-memory caching option. Unlike Redis, it does not persist the data in a data store. It also gives high performance, but it loses data on server restarts. It has less fault tolerance when compared with Redis.

**2.3.2.3 Data store**: Persisting the view state, session state, and application state can use many server-side storage options. The choice of storage depends on the latency and scale requirements of an application. Hence, the pros and cons of each storage option should be evaluated for determining a storage that suits an application requirements. Some options include:
  A. SQL databases - It provides strong ACID properties. However, it is less performant for highly scalable applications.
  B. Document DB such as MongoDB - This is a NoSQL option which does not offer ACID properties compliance. However, it is highly performant and can handle large scale data. This should be used for structured or semi-structured data.
  C. Key-Value DB such as DynamoDB - This is also a NoSQL option which does not offer ACID properties compliance. It is highly performant and can handle large scale data. This should be used when the data stored is in the form of {key, value} pairs.

**2.3.3 Review of Server Side State Management:**
- View State technique is simple to implement.
  - It does not require any server-side configuration.
  - The state is automatically sent back and forth between HTTP requests.
  - It is limited to a single web page, and not shared across pages.
  - It can significantly increase the size of HTML pages, leading to slower load times and higher bandwidth usage. This makes it not suitable for large amounts of data.
- Session State technique stores user-specific data across multiple pages and requests.
  - It is effective for storing user authentication, preferences, and transient data like shopping carts.
  - It helps create personalized experience and improves user satisfaction by preventing unnecessary relogins.
  - It requires careful management to avoid data loss or security issues, such as session hijacking.
  - Persisting session data in a state server or database can add complexity and overhead to the application.
- Application State technique is suitable for storing data that is common to all users and needs to be accessed across the entire application lifecycle.
  - It is useful for configuration settings, health metrics, and other shared data.
  - It helps persisting data to avoid loss across server restarts and in multi-server environments.
  - It does not inherently provide user-specific data isolation, which can lead to conflicts in multi-user scenarios
  - It requires additional mechanisms to persist data across server restarts and ensure data consistency in multi-server deployments.

# Conclusion:

Application State Management (ASM) has become a critical aspect of modern web and mobile application development, addressing the need for consistency, performance, and user-friendliness. This review examined various ASM techniques across four main categories: Local State Management, State Management Libraries, Server-Side State Management, and Local Storage Solutions.

Key findings include:

1. Local State Management techniques in popular frameworks like React, Angular, and Vue offer efficient handling of component-specific data, improving performance for use cases like form handling and presentation logic changes.
2. State Management Libraries such as Redux, Recoil, Zustand, MobX, NgRx, and Vuex provide robust solutions for managing global state and cross-component communication. Each library has its strengths, with Redux being widely adopted due to its battle-tested nature, while newer libraries like Zustand and Recoil offer simplified APIs and fine-grained control respectively.
3. Server-Side State Management techniques, including View State, Session State, and Application State, play crucial roles in maintaining consistency between client and server, improving latency, and enhancing user experience. Technologies like Blazor Server, caching solutions (Redis, Memcache), and various data stores offer different approaches based on application requirements.
4. The choice of ASM technique depends heavily on the specific needs of the application, considering factors such as scale, performance requirements, and development team expertise.
5. Certain use cases can be achieved at both server side and client side. Choice to use a particular technique depends on security needs and available infrastructure for the application. More sensitive applications which deal with PII data of users should use server side ASM, for example a banking application while others may implement ASM using popular client side techniques, for example an e-commerce cart management.
6. Certain technologies are more widely adopted and thus more battle proven. For example Redux for client side ASM and Redis for server side ASM.

When selecting an ASM technique with its technology open challenges described below should be considered.

1. Complexity vs Simplicity: Balancing the need for powerful state management capabilities with the desire for simpler, more intuitive APIs remains an ongoing challenge. While libraries like Redux offer robust solutions, they often come with a steeper learning curve and more boilerplate code.
2. Performance at Scale: As applications grow in complexity and size, maintaining performance while managing increasingly complex state becomes challenging. Optimizing state updates and minimizing unnecessary re-renders are areas that require ongoing attention.

3. Synchronization Between Client and Server: Ensuring seamless synchronization of state between client-side and server-side, especially in real-time applications or those with offline capabilities, presents significant challenges.
4. Security Concerns: As more state is managed and potentially persisted, ensuring the security of sensitive user data becomes increasingly important. This includes both client-side storage and server-side caching mechanisms.
5. Cross-Platform Consistency: With the rise of cross-platform development, maintaining consistent state management across web, mobile, and desktop versions of an application poses unique challenges.
6. Testing and Debugging: As state management becomes more complex, developing effective testing strategies and debugging tools for state-related issues becomes more challenging.
7. Adapting to Emerging Paradigms: As new programming paradigms and architectural patterns emerge (e.g., micro-frontends, server components), state management techniques will need to evolve to accommodate these new approaches.
8. Optimizing for Edge Computing: With the growing trend towards edge computing, adapting state management techniques to work efficiently in distributed environments close to the user presents new challenges and opportunities.
9. Industry Adoption: When choosing a particular technology and technique of ASM, community support and market adoption should be taken in consideration along with complexity of implementation.

Addressing these challenges will be crucial for the continued evolution of Application State Management, ensuring that developers can build increasingly sophisticated applications while maintaining performance, security, and user experience.